\documentstyle[aps,floats]{revtex}

\begin{document} 

\title{Complexity spectrum of some discrete dynamical systems}

\author{N. Abarenkova}
\address{Centre de Recherches sur les Tr\`es Basses Temp\'eratures,
B.P. 166, F-38042 Grenoble, France\\
Theoretical Physics Department, Sankt Petersburg State University,
Ulyanovskaya 1, 198904 Sankt Petersburg, Russia}

\author{J.-Ch. Angl\`es d'Auriac}
\address{Centre de Recherches sur les Tr\`es Basses Temp\'eratures,
B.P. 166, F-38042 Grenoble, France}

\author{S. Boukraa}
\address{Institut d'A\'eronautique, Universit\'e de Blida, BP 270,
Blida, Algeria}

\author{J.-M. Maillard}
\address{ LPTHE, Tour 16, 1er \'etage, 4 Place Jussieu, 
75252 Paris Cedex, France}

\date{\today}

\maketitle

\begin{abstract} 
We first study birational mappings generated by the composition 
of the matrix inversion and of a permutation 
of the entries of $\, 3 \times 3\,$ 
matrices. We introduce a semi-numerical analysis
which  enables to compute the Arnold complexities 
for all the $9!$ possible birational
transformations. These complexities correspond to
a spectrum of eighteen algebraic values. We then drastically 
generalize  these results, 
replacing permutations of the entries by
homogeneous polynomial transformations of the entries possibly 
depending  on many parameters.
Again it is shown that the associated birational,
or even rational, transformations yield algebraic values
for their complexities.
\end{abstract}

\vskip .2cm 

\vskip .2cm 


\vskip .2cm 


\section{Introduction and recalls}
Birational transformations have been seen to be a powerful
tool to analyze the symmetries of the parameter space of
lattice models of statistical mechanics~\cite{BoMaRo95,BoMa95} 
and to seek for some possible Yang-Baxter 
integrability~\cite{bmv2,bmv2b}. Beyond the lattice 
statistical mechanics framework, birational transformations
are worthy to be studied {\it per se}, as discrete dynamical systems.
Discrete dynamical systems have been intensively
studied (see for example~\cite{Ott,ASY}).
 Among them polynomial examples, like the Henon map~\cite{H76},
have been precious to understand some features of chaos.
Beyond, rational mappings are of special interest since they allow
some analytical calculations.  Furthermore,
the rational transformations also allow 
numerical calculations which can be performed with 
any wanted precision : the existence of singularities
in the rational transformations one iterates, 
and their possible ``proliferation'' is not in fact 
a numerical obstruction.
We will first consider
mappings generated by the composition of the matrix inverse
and some arbitrary, but fixed, permutation of the entries
of $q \times q$ matrices. The results, displayed in this paper,
are given for $q=3$, but are actually valid, {\it mutatis mutandis}
\cite{BoMa95}, for arbitrary $q$ values.

\subsection{Recalling a previous $3 \times 3$ analysis}
Integrability of a mapping amounts to saying that all the orbits
of the iteration correspond to elliptic, or rational,
algebraic curves~\cite{BoMaRo93a,BoMaRo93b}. From the point of view 
of the growth\footnote{When one iterates a rational transformations 
the ``size'' of the successive rational expressions, corresponding to
the $N$-th iterate, grows, in general, exponentially. In particular
the {\em degree} of these successive rational expressions
has, generically, an exponential growth~\cite{BoMaRo95,HiVi97}. Growth
of the calculations related with factorizations were also
introduced by Veselov for some particular Cremona 
transformations~\cite{Ve89,MoVe91,Ve92}.
} of the complexity
of the successive iterations~\cite{BoMaRo93a,BoMaRo93b}, such 
integrability in curves
always yields a {\em polynomial growth} of the
 calculations~\cite{BoMaRo95,BoMa95},
instead of the exponential growth one generically 
expects. Conversely polynomial
growth is not restricted to integrability in curves but
may correspond to orbits ``densifying'' Abelian 
varieties~\cite{BoMaRo95,BoMa95}.

A first exhaustive analysis of all the 
$\, 9!\, $ birational transformations generated
by the composition of the matrix inversion and of a (fixed) permutation
of the entries of $3 \times 3$ matrices has  already been performed
concentrating on the extraction of integrable 
mappings~\cite{Zittartz}. This analysis was exhaustive, but
restricted to particular integrability 
criteria\footnote{Associated with particular
 recursions~\cite{Zittartz} on some 
``determinantal'' variables~\cite{BoMaRo93a,BoMaRo93b}
 $\, x_n$'s, also introduced here in section (\ref{factgenfunc}).}.
Even from this ``integrability-digger'' point of view
some integrable mappings are missing (for example
the so-called ``Class III'' mappings of~\cite{BoMaRo93c},
as well as some polynomial growth situations).
In the first part of this paper we will revisit these
$9!\, =\, 362880$ birational mappings 
without any {\it a priori} integrability criterion
and with the help of a new equivalence relation 
among permutations (symmetry). This analysis exactly yields
{\em all} the polynomial growth situations, 
and, far beyond, {\em classifies} the exponential growth situations.
The classification relies on the value of the
{\em Arnold complexity}~\cite{A} of the mapping. This complexity
can be obtained~\cite{zeta} from generating functions associated
with factorization schemes~\cite{BoMa95} detailed below.

\subsection{Factorization scheme and generating functions}
\label{factgenfunc}

We use the same
 notations as in~\cite{BoMaRo93a,BoMaRo93b,BoMaRo93c}, that is, we
introduce the following two transformations, the usual matrix
inverse $\widehat{I}$
and  the {\em homogeneous} matrix inverse $I$ :
\begin{eqnarray}
\widehat{I}:\, \, \, \, \,  M_0 \, \longrightarrow \, \, M_0^{-1} \, ,
 \qquad \qquad
\hbox{and :}\qquad \qquad 
I: \, \, \, \, \,  M_0 \, \longrightarrow \, \, \, \det(M_0) \cdot M_0^{-1}
\end{eqnarray}
The homogeneous inverse $I$ is a homogeneous
 polynomial transformation,
which associates, with each entry of $\, M_0 $,  its 
corresponding cofactor. 
Transformation $t$ is any (fixed) permutation
of the entries of the $\, 3 \times 3$
matrix. We also
introduce
 the (generically infinite order) transformations :
\begin{eqnarray}
\label{defK}
K\, =\, t \cdot I \qquad \quad \hbox{and} \quad  \qquad
\widehat{K}\, =\, \, t \cdot \widehat{I}
\end{eqnarray}
Transformation $\widehat{K}$ is clearly
a {\em birational transformation}~\cite{BoMaRo93a,BoMaRo93b}.

For all the various birational transformations
associated with permutations of the entries of $\, 3 \times 3$ matrices,
the following factorization relations
happen to occur {\em at each} iteration step~\cite{BoMa95} :
\begin{eqnarray}
\label{rappel}
&&f_1\, = \, det(M_0) \, , \quad 
M_1\,= \, K(M_0) \,, \quad
f_2\  = \, \frac{det(M_1)}{f_1^{\phi_1}} \, , \quad
M_2\,= \, \frac{ K(M_1) }{ f_1^{\eta_0}} \, , \quad
f_3\, = \, \frac{det(M_2)}{f_1^{\phi_2} \cdot f_2^{\phi_1} } \, ,\quad 
M_3\,= \, \frac{K(M_2)}{f_1^{\eta_1} \cdot f_2^{\eta_0}} \,, \nonumber
\end{eqnarray}
and for arbitrary $n$ :
\begin{eqnarray}
\label{detgen}
&&\det(M_n) \, \, \, = \, \, \,  \, \,  f_{n+1} 
\cdot f_{n}^{\phi_1} \cdot f_{n-1}^{\phi_2} 
 \cdot f_{n-2}^{\phi_3}  \cdot f_{n-3}^{\phi_4} 
 \cdot f_{n-4}^{\phi_5} \,  \cdots \,  f_{1}^{\phi_n} \\
\label{Kgen}
&&K(M_n) \, \, \, =  \, \, \, \,  \, M_{n+1} 
\cdot f_{n}^{\eta_0} \cdot f_{n-1}^{\eta_1} 
 \cdot f_{n-2}^{\eta_2}  \cdot f_{n-3}^{\eta_3} 
 \cdot f_{n-4}^{\eta_4} \,   \cdots \,  f_{1}^{\eta_{n-1}} \\
\label{Kdetgen}
&&\det(M_n) \cdot M_{n+1} \,  \, \, = \, \,  \,
 \,  \, \,  \,  \bigl( f_{n+1}^{\rho_0} \cdot f_{n}^{\rho_1} 
 \cdot f_{n-1}^{\rho_2}  \cdot f_{n-2}^{\rho_3}  \cdot
 f_{n-3}^{\rho_4} \,  \cdots \,   f_{1}^{\rho_{n}}
\bigr) \cdot K(M_n)
\end{eqnarray}
defining the positive integer exponents 
$\eta_n$, $\, \phi_n \,$ and $\rho_n$. 
The $\, f_n$'s are homogeneous polynomials of the entries of $\,
M_0$. These
 factorizations allow to define, at each iteration step,
the successive $f_n$'s one can ``factor out'',
and the ``reduced matrices'' $M_n$'s, such that their entries are
homogeneous polynomial expressions of the initial entries, and 
have no further factorization.
One finds out, looking at the first
thirties iterations, that one recovers 
the {\em same} exponents ($\eta_n$, $\phi_n$, $\rho_n$)
at each iteration step (up to the last emerging coefficient
for $f_1$). We assume that this regularity property holds 
for arbitrary $n$. This regularity\footnote{In fact it is shown
in~\cite{zeta} that other slightly more general factorizations
scheme can occur on some $\widehat{K}$-invariant
subvarieties (yielding smaller Arnold complexity values). Such 
slightly more general factorizations
scheme will also be detailed below (see Appendix B). However for the
 transformations $\, K$ associated with permutations of $q \times q$ 
matrices, for a generic initial matrix,
 one gets factorization schemes like (\ref{detgen}),
(\ref{Kgen}),
also depicted in~\cite{BoMa95}.}
 property assumption is crucial in our analysis.

We will denote $\alpha_n$ the degree of the determinant of
 matrix $M_n$, and $\beta_n$ the degree of 
 polynomial $\, f_n$ and  $\alpha(x)$, $\beta(x)$, $\eta(x)$, $\phi(x)$ 
and $\rho(x)$,  the generating functions of the degrees
 $\alpha_n$'s, $\beta_n$'s, and of the
 exponents $\eta_n$'s, $\rho_n$'s
 and $\phi_n$'s 
in the factorization schemes :
\begin{eqnarray}
\label{g}
\alpha(x)\, = \,  \sum^\infty_{n=0} \alpha_n \cdot x^n, \;\; \, \, 
\beta(x) \, = \,  \sum^\infty_{n=1} \beta_n \cdot x^n, \;\; \, \,
\eta(x) \, = \,  \sum^\infty_{n=0} \eta_n \cdot x^n, \;\; \, \,
\phi(x) \, = \,  \sum^\infty_{n=1} \phi_n \cdot x^n, \;\; \, \,
\rho(x) \, = \,  \sum^\infty_{n=0} \rho_n \cdot x^n 
\nonumber 
\end{eqnarray}
where $\, \alpha_0 \, = \, 3 $ and  $\, \beta_1 \, = \, 3 $.
It is straightforward
to show~\cite{BoMa95}  that the existence
of the {\em stable} factorization scheme (\ref{detgen}),
(\ref{Kgen}) yields the following simple
 linear relations between these various ``degree generating functions'' and 
``exponents generating functions'' :
\begin{eqnarray}
\label{etax}
&&\alpha(x)\, + \, 3\cdot  x \cdot \eta(x) \cdot \beta(x) \,\,\,  = \,\,
\,  3\, + \, 2 \cdot x\, \cdot \alpha(x)  \\
\label{defzet}
&&x \cdot  \alpha(x) \, \, = \,\,\,\, \,\, \phi(x) \cdot \beta(x)  \\
\label{rhox}
&&3 \,\, + \, 3 \cdot \rho(x) \cdot  \beta(x) \,\,\,\, \,
 =\,\,\,\,\, \,  (1+x)\cdot \alpha(x)
\end{eqnarray}
When analytically  iterating an arbitrary transformation
$K$, the degree of the successive polynomial expressions
one encounters, grow exponentially : $\alpha_n$ or $\beta_n \, \simeq
\, \lambda^n$.
 $\lambda\, $ measures the grows of the calculations 
and  identifies
with the notion of Arnold complexity~\cite{A,zeta}. From now on
$\, \lambda \, $ will be called the {\em complexity}.
When the degree generating functions $\alpha(x)$ or $\beta(x)$
happen to be rational functions, the complexity $\, \lambda\, $ is
obviously the inverse of the pole of smallest modulus. Recalling 
the ``determinantal'' variables~\cite{BoMaRo93a,BoMaRo93b,BoMaRo93c}
$\, \, x_n$'s  defined by :
\begin{eqnarray}
\label{defxn}
x_n(M_0) \, = \,\, \, \det(\widehat{K}^{n+1}(M_0)) \cdot \det(\widehat{K}^{n}(M_0))
\end{eqnarray}
one finds out that these determinantal variables happen to decompose
on a product of the homogeneous polynomials $\, f_n$'s {\em only} :
\begin{eqnarray}
\label{detoverdetxn}
x_n(M_0)\, \, \, =\, \, \,\, \, \, \,  f_{n+1}^{w_0} 
\cdot f_n^{w_1} \cdot f_{n-1}^{w_2} 
\cdot f_{n-2}^{w_3} \cdot f_{n-3}^{w_4}
 \cdot f_{n-4}^{w_5} \, \,  \cdots
\, \, \,
\end{eqnarray}
which defines some, at first sight,
``new'' exponents\footnote{The $\, w_n$'s are {\em  relative}
integers and not natural integers like exponents $\eta_n$, $\phi_n$
and $\rho_n$.
}  $\, w_n$'s and consequently 
a, at first sight,
``new'' generating function $ {\cal W}(x)$ :
\begin{eqnarray}
\label{defcalW}
{\cal W}(x)  \, \, \, = \, \, \, \, \,
 \,\sum^\infty_{n=0} w_n \cdot x^n  \;\; \, \, 
\end{eqnarray}
It is worth noticing that the determinantal variables 
$x_n$'s induce the homogeneous polynomials
$\, f_n$'s emerging from the factorization schemes (\ref{detgen}),  (\ref{Kgen}) 
and no other homogeneous polynomials.
The variables $ \, x_n$'s are
well-suited since they are invariant 
under a multiplication
of  $ \, M_0$ by a constant :  
$ \, M_0\, \,  \rightarrow \, \,   {\rm Cst} \cdot M_0$.
In other words the $x_n$'s are homogeneous expressions of degree zero.
Concentrating on the degrees of the left-hand side,
and right-hand side, of (\ref{detoverdetxn}), 
one gets the following ``degree equation'' :
\begin{eqnarray}
0 \, \, = \, \, \, \,  \beta_{n+1} \cdot {w_0} \, + \, \,  
 \beta_{n} \cdot {w_1}  \,
 \,  \,   + \, \cdots  \, 
  +\, \beta_{n-p} \cdot {w_{p+1}} \, + \,    \cdots\,
 + \beta_1 \cdot {w_{n}} 
\end{eqnarray}
from which one immediately deduces the simple functional equation :
\begin{eqnarray}
\label{Wb}
{\cal W}(x) \cdot \beta(x) \, \, = \, \, \, \, 
\beta_1 \cdot w_0\, \, = \, \, \, \,3 \cdot x 
\end{eqnarray}
This result is immediately generalized to $\, q \times q\, $
matrices. Relation (\ref{Wb}) becomes $\, {\cal W}(x) \cdot \beta(x)
\, \, = \, \,
q \cdot x$.
From (\ref{Wb}) one  actually sees that  ${\cal W}(x)$ is 
{\em not} a new generating function : it is simply related
to the degree generating function $ \,\beta(x)$.
The complexity $\lambda$ is associated to the {\em zeroes}
of  $\, \, {\cal W}(x)$.

From  relations~(\ref{etax}), (\ref{defzet}) and (\ref{rhox}),
 one easily gets the ``degree generating functions'' 
$ \,\alpha(x)$ and $ \,\beta(x)$ from two of the
``exponent generating functions'' (for instance $\, \phi(x)\, $
and  $\, \eta(x) $ or $\, \eta(x) $ and $\, \rho(x)$).
As a matter of fact,  for most of the 
permutations, the factorization schemes are {\em periodic}
($\eta_n\, = \, \eta_{n+N}$,
 $\phi_n\, = \, \phi_{n+N}$  and  $\rho_n\, 
= \, \rho_{n+N}$ for some integer $N$).
Consequently, the exponent generating functions $\, \phi(x)\, $
and  $\, \eta(x)\, $, or $\, \rho(x)$, are  {\em rational} functions
with $N$-th root of unity poles~\cite{BoMa95} (see, 
for instance, the exponent generating function $\,
 \rho(x)\, $ in~(\ref{root1}) or (\ref{1x6})). In a second 
step one deduces,
from relations (\ref{etax}), or (\ref{defzet}),  
rational expressions for the degree generating functions 
$\alpha(x)$ and $\beta(x)$.
However it will be seen below that, for some permutations, 
the factorization schemes {\em are still regular}, but not
with periodic exponents (see the exponent generating function $\,
 \rho(x)\, $ in~(\ref{m063}) or (\ref{mumu1}))
: the exponents $\eta_n $,  $\phi_n$ and $\rho_n$ 
{\em grow exponentially}, but one remarks that the associated 
generating functions $\eta(x)$, $\phi(x)$, $\rho(x)$
are {\em still rational}, and thus $\alpha(x)$ and $\beta(x)$
{\em are also rational}.  
The exponent generating functions can be 
seen as an ``encoding'' of the
degree generating functions, and thus of the complexity $\,\, \lambda$.
Remark that all these 
rational expressions involve {\em integer} coefficients,
yielding {\em algebraic} values for their poles
and for the growth of the calculations :
the degrees of the successive rational expressions, namely
 $\alpha_n$'s and $\beta_n$'s 
grow like $\lambda^n$, where  $\lambda\, $ 
is an {\em algebraic number}, and for regular factorization schemes
(like (\ref{m063}) or (\ref{mumu1})), see below), the exponents
 $\, \eta_n \, $
and $\, \phi_n \, $ grow like $\, \mu^n$, where $\, \mu \, $ is 
the ``scheme complexity''. Note that $\, \mu \, $
 is obviously such that $\, \mu \le \lambda$. 
Exponent $\mu$ is also the inverse
of the pole of smallest modulus of the 
exponent generating 
functions. Complexity  $\lambda$ 
 allows all kinds of handy, efficient, and 
formal, or semi-numerical, calculations.
We will present below such a semi-numerical method 
and apply it to {\em all the} permutations
of entries of $\, 3 \times 3$ matrices\footnote{However one should
 keep in mind that there is nothing specific with  $\, 3 \times 3$
matrices. These results simply generalize to  $\, q \times q$ matrices
(see for instance~\cite{BoMa95}).}.

\section{Complexity spectrum analysis for permutations}
\subsection{A semi-numerical method}

All these considerations allow us
 to design a {\em semi-numerical method} to get
the value of the complexity $\, \lambda$ for the iteration
of {\em rational} transformations.
The idea is to iterate, 
with $\widehat{K}$, a generic
 initial matrix with {\em integer} 
entries. After one iteration step the entries become rational
 and we follow the magnitude
of the successive numerators and denominators. During the first 
few iteration steps some ``accidental''
 simplifications may occur,
but, after this transient regime, the integer denominators (for instance) grow
like $\lambda^n$,  where $n$ is the number of iterations.
One can systematically improve the method as follows : 
the initial matrix is chosen in such a way 
that it avoids, as much as possible,
any ``accidental'' additional factorization in comparison with
the factorization scheme associated with a generic matrix.
For instance, in a factorization scheme framework
like (\ref{detgen}), (\ref{Kgen}), one chooses 
the initial matrix $\, M_0\, $ 
with integer entries such that the determinant, 
and most of its cofactors, are prime numbers as large as possible.
One may impose further constraints on the initial matrices, 
for instance, 
that the first homogeneous polynomials $f_2$ and $f_3$
are also,  as large as possible, prime numbers.
These conditions down-size the probability that all the entries
of the reduced matrices $\, M_n\, $, or  the polynomials $f_n$'s, 
could be divisible by some accidental additional $\, f_1$, $f_2$ or $f_3$.
Such initial matrices, well-suited for the iteration of
the homogeneous transformation $K$, are also 
well-suited 
for the iteration of the (bi-)rational 
transformation $\widehat{K}$. In practice we start with a set of initial matrices 
and keep only the
one for which the {\em less} factorizations occur (non-generic
factorization can only correspond to
{\em additional} factorizations).

The computations are done using an infinite precision
C-library~\cite{C}. We perform as many
iterations as possible during a given CPU time $\, T$.
This number of iterations, $n$, is such that $\, T \, \simeq \,
\lambda^n$. For $\, \lambda \, $ close to $\, 2$ and $\, T \, = \, $
60 seconds, $\, n$ is of the order of twenty and 
a best  fit of the logarithm of the numerator as a linear function
of $n$, between $\, n=10$ and $n=20$, gives the value of $\, \lambda\, $ within
an accuracy of $0.1\%$. For smaller values of $\, \lambda\, $ 
(typically $\lambda < 1.5$) the
number of iterations is larger, but the accuracy, for a given CPU time,
is smaller. In such ``difficult'' cases one analytically finds
the factorizations up to $\, n\, = \, 7$
and implement the first steps
of these factorization schemes in the semi-numerical method. 
We are then almost guaranteed that no accidental factorizations
will occur for $\, n > 7$, and therefore we can average over 
many initial matrices. Even so it remains difficult to discriminate
between
a truly polynomial growth~\cite{BoMaRo95,BoMa95} ($\lambda \, = \, 1$) and 
an exponential growth with $\, \lambda \, \simeq 1$.  The complexity values
close to one clearly need to be revisited 
by other methods we present below.

\subsection{Equivalence relations between permutations}

Even if this semi-numerical algorithm is efficient
it is quite time consuming to use it directly
on the  $\, 9!$ permutations. 
To classify the complexities associated to a large 
set of (birational) transformations like the one
associated to the $\, 9!$ permutations of $\, 3 \times 3\, $
matrices, one certainly needs to reduce
this set as much as possible. For instance one can try to find
symmetries such that two permutations, related by the symmetry,
yield the same complexity $\, \lambda$.
These symmetries allow to build {\em equivalence classes}
and, thus, to restrict the exhaustive analysis to
a only one representent
in each class.  Furthermore 
one may have the prejudice that any non trivial symmetry 
could enable to explain a possible
integrability structure of the mappings
and beyond, the structures associated with the
classification
of the Arnold complexity of these mappings.

There actually exist quite trivial symmetries,
corresponding to relabeling of rows and columns~\cite{Zittartz},
for which the complexities of the associated $\, K$'s
are obviously equal.
It is possible to go a step further and define
a set of equivalence relations ${\cal R}^{(n)}$
between the permutations, yielding new 
equivalence classes such that any two permutations
in the same ``new'' equivalence class, $\, {\cal R}^{(n)}$, 
automatically have the {\em same complexity} $\lambda$.
Equivalence relation ${\cal R}^{(n)}$ amounts to saying
that two equivalent permutations 
are such that the $n$-th power of their associated transformations
$\, \widehat{K}\, $ 
 are conjugated (via particular permutations,
product of row permutations, column permutations
and possibly the transposition, see appendix A for more details).
An exhaustive inspection has shown 
 that the equivalence relations  ${\cal R}^{(n)}$'s
``saturate'' after $n\, =\, 24$: with obvious notations
  $\, {\cal R}^{(\infty)}\, = \,{\cal R}^{(24)}$. One finds out that 
the ``ultimate''  $\, {\cal R}^{(\infty)}\,$
equivalence classes {\em can only have} $72$, or $\, 144$, elements.
Among the ``ultimate''  $\, {\cal R}^{(\infty)}\,  $
classes, one wants to distinguish between the classes
that were already  ${\cal R}^{(1)}\, $ classes, that we will
denote from now on ${\cal R}^{(1)}_{72} $, or ${\cal R}^{(1)}_{144}$,
according to their number of elements , and the other ones
we denote $\, {\cal R}^{(\infty)}_{72}\,$
 or $\, {\cal R}^{(\infty)}_{144}$. Being an 
$\,{\cal R}^{(\infty)}\, $ equivalence class which does not reduce to
a $\, {\cal R}^{(1)}\, $ equivalence class, means the existence of
several non trivial relations between the permutations in the 
$\,{\cal R}^{(\infty)}\, $ equivalence class (see (\ref{fund}) in
appendix A). This implies strong constraints on the respective
orbits.
One thus expects more properties, and structures, inherited from this fact.
The $\, 362880 \, $ permutations are  grouped
into $ \, 2880\, $   equivalence classes (instead of 
$ 30462 \, $ ``relabeling''
 equivalence classes in~\cite{Zittartz}). 
In Tab.~\ref{latable}  the number of the respective
 ${\cal R}^{(1)}_{72} $, ${\cal R}^{(1)}_{144}$,
 $\, {\cal R}^{(\infty)}_{72}\,$ or
 $\, {\cal R}^{(\infty)}_{144}\,$ classes is displayed.
Since the complexities do not depend on the  chosen representent, we
picked a representent in each $\, \cal{R}^{(\infty)}$  class 
 and performed, for it, the semi-numerical method
previously explained. 

 For $\, 3 \times 3\, $ matrices, the complexities are necessarily 
such that : $\,\, 2 \,\ge\, \lambda\, \ge\, 1\, $. Remarkably,
 instead of getting a quite complicated 
distribution, or spectrum, of values for the complexities,
 we have  obtained values which are always very close,
up to the accuracy of the method, to a set 
of seventeen values given in the left column of Tab.~\ref{latable}
(see below) and, of course, the integrable value $\, \lambda \, = \, 1$.
To test the accuracy of the method we  
got complexities for two representants of
the {\em same } class (that should, as we know, 
have the same complexity value exactly).
We always obtained an equality
of the corresponding complexities, up to an error
of $10^{-3}$. This accuracy is however not always sufficient
 enough to discriminate between some 
complexities displayed in the left column of Tab.~\ref{latable}.
In order to fix our mind it is necessary to obtain the exact expressions
of these complexity values, for instance by getting the 
factorization scheme (\ref{detgen}), (\ref{Kgen}),
 and thus the generating functions
$\alpha(x)$ and $\beta(x)$. 

\subsection{Revisiting the complexity 
spectrum via exact factorization schemes}
\label{revisit}
For most of the  $\, {\cal R}^{(\infty)}\,$ equivalence
classes ($2832$ out of $2880$), the complexity
 values, obtained with our semi-numerical
method, are extremely close to the upper limit
$\lambda =\, 2\,$. In fact one can figure out that 
these complexity values are actually exactly equal to $\, 2$.
Therefore we can focus on the analysis of the remaining
$48$ classes, finding systematically 
their factorization schemes and associated
generating functions. We actually found these factorization schemes 
and the  associated
generating functions, and were actually able to see
that the previous numerical spectrum  {\em exactly} 
corresponds to  eighteen
algebraic values listed in Tab.~\ref{latable}. Among these eighteen
algebraic values, let us take four 
illustrative examples. We give for each example, the permutation
representing
the  $\, {\cal R}^{(\infty)}\,$ equivalence
class, 
the value of $\, \lambda\, $, and $\, \mu $, defined 
in section~\ref{factgenfunc}, the expressions 
of  $\, \beta(x)$ and $\rho(x)$, since they respectively correspond to
 the simplest ``degree generating function''
and ``exponent  generating function''.  The other 
generating functions can be deduced from 
these two, using linear functional relations (\ref{etax}), (\ref{defzet})
and  (\ref{rhox})
between the generating functions~\cite{BoMa95}. 
Furthermore relation (\ref{detoverdetxn}) 
remains valid for all the factorization schemes 
associated with all the various permutations studied here.
We first give the permutation $\, t\, $ itself, using the 
notation, already used in~\cite{Zittartz},
where $\, p_0p_1p_2p_3p_4p_5p_6p_7p_8\, $ means that $\, (t
\tilde{M})_i \, =\, \tilde{M}_{p_i}$,
the entries of the matrix being enumerated consecutively
(i.e. $ M_{11} = \tilde{M}_0, \, \, \, M_{12} = \tilde{M}_1, \, \, \, M_{13} = \tilde{M}_2, $
$\, M_{21} = \tilde{M}_3, \, \, \cdots , \, \, M_{33}\, =\, \tilde{M}_8$).
\vskip .3cm 
\noindent
$\bullet$ First example.    
Permutation $\, 407326518\,$ yields  
$\lambda \simeq  \, 1.61803 \cdots \, $   and $ \, \, \mu \, = \, 1$
and :
\begin{eqnarray}
\label{root1}
\qquad {{ \beta(x) } \over {3 \, x }} \, =
 \, \,  {{ 1-x^2} \over {1-x-x^2}} \,,  \qquad  \quad \qquad
 \rho(x) \, = \, \,{\frac {1}{ (1-\, x )^{2}\cdot (1+x )}}
\end{eqnarray}
$\bullet$ Second example. Permutation $\,\,417063582 \,$
yields $\lambda \simeq \, 1.83928 \cdots \,$
 and $\, \, \mu \, \simeq \,
1.32471 \cdots \, $ and :
\begin{eqnarray}
\label{m063}
\qquad  {{ \beta(x) } \over {3 \, x }} \,
 \, = \, \,{\frac { 1\, -\, x^2\, -x^3}{
(1-x )^{2}
\cdot (1+x)\cdot (1\, -x\, -\, x^2 \, -\, x^3 )}}
 \, , \qquad  \rho(x) \, = \,\,
 \, {\frac { (1 \, + \, x ) \cdot (1\, -\, x \, + \, x^4
)}{1\, -x^2\, -\, x^3}}
\end{eqnarray}
$\bullet$  Third example. Permutation $\,\, 164273085 \,$ yields 
$\, \lambda \simeq \, 1.83928 \cdots \, \, $ and $\, \, \mu \, = \, 1$ and :
\begin{eqnarray}
\label{1x6}
\qquad {{ \beta(x) } \over {3 \, x }} \, = \, \,
 {\frac {   1\, +x\, +x^2}{1-x-x^2-x^3}}
 \, , \qquad \qquad \rho(x) \,  \,=
\, \, \, {{1+x^3+x^4+x^5} \over {1-x^6}} 
\end{eqnarray}
$\bullet$ Fourth example. Permutation $\,174528603 \,$ yields 
$\lambda \simeq \,  1.97458 \cdots  \, $ and $\, \mu \, \simeq \,
1.32471 \cdots \, $ and :
\begin{eqnarray}
\label{mumu1}
 \quad {{ \beta(x) } \over {3 \, x }} \,
 \, = \,\,
{\frac {1\, -{x}^2\, -{x}^{3}}{ (1-x ) \cdot (
1 -x-2\,{x}^2-{x}^3\, +{x}^{4} \,+2\, x^5\, +{x}^6 )}}
 \, ,   \quad \rho(x) \, = \, \,
{{ 1-x+x^7+x^8 } \over {(1-x+x^2) \cdot (1-x^2-x^3)}}\,
\end{eqnarray}

The exhaustive analysis of  the 
factorization schemes, and the associated 
degree, and exponent, generating 
functions ($\alpha(x)$, $\beta(x)$,
$\eta(x)$, $\phi(x)$ and $\rho(x)$), 
confirms that the complexities
are actually independent of the representent in the equivalence 
class. On the contrary, the
 factorization schemes and the associated 
degree, and exponent, generating functions may 
depend\footnote{Considering one $\, \cal{R}^{(\infty)}\, $ equivalence class,
one does not get as many  factorization schemes as the number
of elements in the  equivalence class. It seems, inspecting directly
all the $\, 9! \, $ factorization schemes (but only up to twelve
iteration steps), that, most of the time,
one gets, 
at most, two possible  factorization schemes for a given 
$\, \cal{R}^{(\infty)} \, $ equivalence class, and that the set of all
the possible factorization schemes would be twenty one (besides
the polynomial growth situations which can be quite ``rich'').
} on the chosen representent in the equivalence class. In other 
words to two permutations in the same 
class of equivalence 
correspond the same (up to $\, \, 1\, -x\, $ or $\, 1\, +x\, $,
 or $N$-th root
of unity factors)
denominators for the degree generating 
functions $\alpha(x)$, $\, \beta(x)$. By contrast the numerators,
as well as the exponents generating functions
are  {\em representent dependent} (see the previous four examples).
Most of time the {\em stability of the factorization scheme}
and thus, in a second step, the occurrence
of {\em rational} generating functions, corresponds to a simple
periodicity of the exponents $\, \eta_n$,  $\phi_n$ or $\rho_n$
in the factorization scheme (\ref{detgen}), (\ref{Kgen}).
This periodicity is simply associated to the fact that the
exponent generating functions have $\, N$-th root of unity poles :
$\, 1-x^2$, $\, 1-x^8 \, $, 
$\,  1-x^6\,,\,  \cdots  \, $ (see $\, \rho(x)\, $ in~(\ref{1x6})).
However one sees, on examples (\ref{m063})
 and  (\ref{mumu1}), that 
one may have a {\em stability} of the factorization scheme
with an {\em exponential growth} of these
 exponents $\, \eta_n$ and $\phi_n$.
These exponent generating functions, of course, have a 
``scheme complexity''  $\mu$ 
 smaller that the growth complexity
 $ \lambda$. 
This ``scheme complexity''  $\mu$ is the inverse of the poles
of $\, \rho(x)$, $\, \phi(x)\, $ or $\, \eta(x)$, that is
 (for (\ref{mumu1})), $\, \mu \, \simeq \,
1.32471 \cdots  \le \lambda \simeq 1.83928 \cdots $. 
Recalling (\ref{1x6}),  for which $\mu\, = \, 1\, $
 and $ \lambda \simeq
1.8392 \cdots $, 
and (\ref{m063}), one sees that 
the {\em same}  growth complexity 
  $ \lambda $ can be 
associated to {\em several}
 ``scheme-complexity'' $\mu$. Conversely,
comparing the fourth example (\ref{mumu1}) 
and the second example  (\ref{m063}), one sees that 
one  ``scheme-complexity'' $\mu$ can actually yield
{\em several} growth  complexities $\lambda$.

\subsection{To sum up}
All these factorization scheme calculations 
confirm the results of the semi-numerical method 
and are 
summarized in Tab.~\ref{latable}.
\begin{table}
\begin{tabular}{||l|l|l|l|l|l|l||} \hline
  $\lambda$ &  Polynomial &  ${\cal R}^{(1)}_{144}$ 
 &$ {\cal R}^{(1)}_{72}$ & 
$ {\cal R}^{(\infty)}_{144}$ &  ${\cal R}^{(\infty)}_{72}$ 
&  Total    \\\cline{1-7}
   Total   &   & 2146   & 660  &  14  & 60   & 2880       \\  \hline \cline{1-7}
  2  &  $1-2\, x$  &  2145 &  640 &  14 &  33 &  2832    \\ \cline{1-7}
  1.97481871 &  $1-2\,x+{x}^{2}-2\,{x}^{3}+{x}^{4}-2\,{x}^{5}+{x}^{6}$ &  0 &  2  &  0 &  0 &  2
  \\ \cline{1-7}
  1.974584654 &  $1-x-2\,{x}^{2}-{x}^{3}+{x}^{4}+2\,{x}^{5}+{x}^{6}$ &  0 &  1  &  0 &  0 &  1
  \\ \cline{1-7}
  1.94893574   & $1-2\,x+{x}^{5}-{x}^{7} $  &  0 &  2
& 0  &  0  &  2      \\  \cline{1-7}
  1.946856268   & $1-x-{x}^{2}-{x}^{3}-{x}^{4}-{x}^{5}+{x}^{6} $  &  0 &  1
& 0  &  0  &  1      \\  \cline{1-7}
  1.93318498 & $1-2\,x+{x}^{4}-{x}^{5}$  &  0 &  1 &  0 &  0 &  1    \\ \cline{1-7}
  1.891103020 & $1-2\,x+{x}^{2}-2\,{x}^{3}+2\,{x}^{4}-2\,{x}^{5}$ 
 &  0 &  0 &  0 &  1 &  1    \\ \cline{1-7}
  1.88320350 & $1-2\,x+{x}^{2}-2\,{x}^{3}+{x}^{4}$  &  0 &  2 &  0 &  6 &  8 
\\ \cline{1-7}
  1.866760399 & $1-2\,x+{x}^{3}-{x}^{4}\,$  &  0 &  1 &  0 &  0 &  1 
\\ \cline{1-7}
  1.860073051 & $1\, - x \, - x^2 -x^4\, - 2 \cdot x^5 \,$  &  0 &  1 &  0 &  0 &  1 
\\ \cline{1-7}
  1.857127516 & $1-2\,x+{x}^{2}-{x}^{3}-{x}^{5}-{x}^{7}+{x}^{8}-2\,{x}^{9}+{x}^{10}$ 
 &  0 &  1 &  0 &  0 &  1 
\\ \cline{1-7}
  1.83928675  & $1-x-{x}^{2}-{x}^{3}$  &  0  &  2  &  0  & 0  &  2     \\ \cline{1-7}
  1.75487766 & $1-2\,x+{x}^{2}-{x}^{3} $ &  1 &  0 &  0 &  0 &  1   \\\cline{1-7}
 1.61803399   & $1-x-{x}^{2}$  &  0   &  3&  0  &  0  &  3  \\ \cline{1-7}
 1.57014731 &  $1-x-{x}^{3}-{x}^{5}$ &  0 &  1 &  0 &  0 &  1  \\ \cline{1-7}
  1.542579599 &  $1-x-{x}^{3}-{x}^{7}-{x}^{8}$ &  0 &  1 &  0 &  0 &  1  \\ \cline{1-7}
 1.46557123  &  $1+x -x^3$ &  0 &  0 &  0 &  2 &  2    \\\cline{1-7}
 1 ( Pol.gr.)    & $ 1-x\,, \,\,\,\, \,\, 1-x^N\, , \cdots $ &  0  &  0 
&  0  &  9  & 9      \\ \cline{1-7}
 1 ( Period.) &    &  0 &  1 &  0 &  9 &  10     \\ \cline{1-7}
 \hline
\end{tabular}
\caption{
\label{latable}}
\end{table}
 Most of the
 $ \, 362880 \, $ birational transformations
considered here do correspond to the
most ``chaotic complexity'', namely the upper bound
 $\, \lambda \, = \, \, 2\, $ : one has $ 359568$
such  $\, \lambda \, = \, \, 2\, $ birational transformations, that is
99.0873 \% of all the birational transformations.
It is known~\cite{Zittartz}, that some symmetry-classes 
correspond to situations where the determinantal
variables
$\, x_n$'s, defined by (\ref{defxn}), are periodic
 (denoted  ``Period.''  in the Table~\ref{latable}).
This $\, x_n\, = \, \, x_{n+N}\, $ situation
may correspond to situations where
 mapping $\, \widehat{K}$, itself, is of finite order (trivial integrability),
but also  to polynomial growth  
situations, that is, $ \,\lambda\, = \, 1 \,$
{\em exactly}. One remarks that  $ {\cal R}^{(\infty)}_{72}$ 
contains {\em all the  integrable},
or {\em polynomial growth},  mappings and, up to one class in 
$ {\cal R}^{(1)}_{72}$, all the  
mappings such that $\, x_n\, = \, x_{n+N}$,
including the situations where 
mapping $\, \widehat{K}$, itself, is of finite order.

\section{Various generalizations}
We now show that all these results also apply for a much larger set of
rational transformations. The number of permutations of  entries
of $\, 3 \times 3\, $ matrices being finite it has been possible to
perform an exhaustive analysis. For more general 
transformations, depending on continuous parameters,
is not anymore possible and we will proceed just with chosen examples.
These examples always combine homogeneous transformations of the
entries of a matrix together with the matrix inversion.
Therefore the transmutation relations detailed in appendix A
still apply, yielding again non-trivial symmetries for these new set 
of transformations.

\subsection{Combining different $K$'s}
\label{mol}
Let us first consider 
 permutation $\, 146237058\, $, and its associated
 $\, \lambda \, \simeq  \, \, 1.97481 \cdots $
transformation $K_1$, and  permutation $\,471562380 \, $ and its 
 $\, \lambda \, \simeq  \, \, 1.54258 \cdots $
transformation $K_2$. Let us compose the two previous
 transformations. From these two ``atoms'' we build the
``molecule'' $ \, {\cal K} \, = \, K_2 \cdot K_1 $. Note that
 $ \, {\cal K} \, = \, K_1 \cdot K_2 $, 
obviously has the same complexity.

This example is an interesting one since
the complexity (obtained from the previous
semi-numerical calculations) 
of the ``molecule'' $ \, {\cal K} \, = \, K_2 \cdot K_1 $ 
is smaller than the product of the two complexities of 
$K_1$ and $K_2$ :  $\, \, \, \lambda( {\cal K}) \, \simeq \,     2.897 \, <
\, 1.9748 \cdot 1.5426 \,  \simeq \,  3.0463\, $.  In general the combination
of two complexities $\lambda_1$ and  $\, \lambda_2$ gives
a complexity for the ``molecules'' larger than the
product $\, \lambda_1 \cdot \lambda_2\, $, often equal to the
upper bound (here $\, \lambda_{\rm upper} \, = \, 4\, $).

The factorization scheme of $ \, {\cal K} \,$ 
is of the same type as the one described in~\cite{zeta},
namely a ``{\em parity-dependent}'' factorization scheme.
It is detailed in appendix (\ref{factomol})
and yields a degree generating function $\beta(x)$ :
\begin{eqnarray}
\label{betamol}
 {{ \beta(x) } \over {3 \, x }} \,
 \, = \,\,\,\, {{ 1+ \, 2\,x\, - \, x^2\, - x^{4} + x^{6} } \over {
1-3\,{x}^{2}+{x}^{4}-{x}^6 -2\,{x}^{8}}}
\end{eqnarray}
The complexity of the molecule $ {\cal K} \,$ 
does not identify with the complexity of $ K_1$, or the one of $K_2$.
It is a true {\em new} algebraic number. This algebraic expression 
for the  complexity of the molecule is in good agreement with the
semi-numerical value obtained above.
We have systematically studied such ``molecules''
for a choice of eighteen representants
of the eighteen complexities of table~\ref{latable}
combined with themselves, and beyond, 
with {\em other} representants. If one barters 
the permutation $\, t_2 \, $ for another 
representent $t_2^{(2)}\, $ in the same ${\cal R}^{(\infty)} \, $ class,
transformation $\, K_2$ being modified accordingly ($ K_2 \, \,  \longrightarrow
\,  \, \,K_2^{(2)}\, $), the new 
``molecule'' $ \, {\cal K}^{(2)} \, = \, \,  K_2^{(2)} \cdot K_1 \, \,
$
yields, in general, {\em another} algebraic value for the complexity
$\, \lambda$ : the equivalence relation ${\cal R}^{(\infty)}\, $
is no longer compatible with the ``molecular structure''.

For all these ``molecules'' the
 {\em parity-dependent} factorization scheme,
yields algebraic numbers for the 
complexities of these molecules in agreement
with the values obtained from our semi-numerical
now applied for the ``molecules''.
Combining among themselves all the permutations yields a large
number of different algebraic complexities, much
larger than the number of complexities obtained combining
only representants of the ${\cal R}^{(\infty)} $ classes among
themselves.

\subsection{From permutations to linear transformations}

We got algebraic results 
on birational mappings associated with permutation of the entries.
We now address the following question :
are these structures (existence of a
stable factorization scheme)
dependent of the fact that we are dealing with permutations ?
In other terms,
does one loose these {\em algebraic} properties when 
deforming the permutations in most general
 transformations ?
The most simple, and natural, 
generalization
amounts to 
replacing the permutation of the entries by {\em linear}
combination on the entries. 

Let us now consider a first  example, namely
 the {\em quite general linear transformation} depending on 
twenty one parameters :
\begin{eqnarray}
\label{linearL}
&&L : \, \quad
 \left [\begin {array}{ccc}
 m_{{1,1}}&m_{{1,2}}&m_{{1,3}}\\
\noalign{\medskip}m_{{2,1}}&m_{{2,2}}&m_{{2,3}}\\
\noalign{\medskip}m_{
{3,1}}&m_{{3,2}}&m_{{3,3}}\end {array}\right ]
\qquad \longrightarrow \qquad \\
&&\left [\begin {array}{ccc}
 m_{{1,1}}&a_{11}\,m_{{1,1}}+a_{12}\,m_{{1,2
}}+a_{13}\,m_{{1,3}}+a_{21}\,m_{{2,1}}+a_{22}\,m_{{2,2}}+a_{23}
\,m_{{2,3}}+a_{31}\,m_{{3,1}}+a_{32}\,m_{{3,2}}+a_{33}\,m_{{3,3}}&m_{{1,3}}\\
\noalign{\medskip}m_{{2,1}}&c_{21}\,m_{{2,1}}\, 
+c_{22}\,m_{{2,2}}+c_{23}\,m_{{2,3}}&m_{{2,3}}\\
\noalign{\medskip}m_{{3,1}}&b_{11}\,m
_{{1,1}}+b_{12}\,m_{{1,2}}+b_{13}\,m_{{1,3}}+b_{21}\,m_{{2,1}}\,
+ b_{22}\,m_{{2,2}}+b_{23}\,m_{{2,3}}+b_{31}\,m_{{3,1}}
+b_{32}\,m_{{3,2}}+b_{33}\,m_{{3,3}}&m_{{3,3}}
\end {array}\right ] \nonumber
\end{eqnarray}
This particular form singles out the rows of the $\, 3 \times 3\, $
matrix (and thus can be understood as an  RCT-compatible form, see
appendix A).
Similarly to the previous paragraphs
let us introduce the homogeneous transformation $\, \,  K \, \, = \, \, L \cdot I\, $.
Factorizations again occur at each iteration step.
These factorizations correspond to a {\em stable factorization scheme 
} giving a growth like $\, \lambda^N$, 
where  $\, \lambda \simeq 1.61803 \cdots $.
It is of the general type described in 
(\ref{detgen}) and (\ref{Kgen}). 
This yields the following generating functions :
\begin{eqnarray}
{{ \beta(x) } \over {3\, x}} \, = \, \, 
\,{\frac {1}{1-x-{x}^{2}}}\, ,\quad  \quad \quad \quad \quad
  \rho(x) \, = \, \,  {{1} \over {1-x}} 
\end{eqnarray}
These results are actually valid for 
any ``sufficiently generic'' choice of the twenty one parameters.
One thus has a first
``universality'' property : the complexity  $\, \lambda\, $
is ``generically'' not dependent of
the previous  twenty one parameters. Furthermore relation (\ref{detoverdetxn}) 
(and consequently relation (\ref{Wb}))
{\em remains also valid} for all the factorization schemes 
associated with all the linear transformations
 studied in this section.
Complexity $\, \lambda \simeq 1.61803 \cdots $ (corresponding to polynomial 
$ 1-x-{x}^{2}$) is a complexity value already
found in table~\ref{latable},  in the sixteenth row.
It is noteworthy that no choice of the twenty one parameters
leads to a permutation of  any of the three
classes corresponding to 
$\, \lambda \simeq 1.61803 \cdots $. Besides the identity, 
the only choice of parameters, leading to a permutation,
is $\, b_{12}\, =\, a_{32}\, =\, 1 $,  all others being zero. 
The permutation is then the 
transposition\footnote{Denoted class IV in~\cite{BoMaRo93c}.} 
$M_{1,2} \, \, \leftrightarrow \, \,M_{3,2}$ 
which corresponds to the complexity 
$\lambda  \simeq 1.46557 \cdots$. This transposition is not isolated
in the twenty-one parameters set of transformations.
Actually if the parameters verify the two conditions 
$\,\,\, b_{12} \cdot a_{32} \, -\,  a_{12} \cdot b_{32}\, \, =\, \, 1\,\,\, $ and
$\,\,\, a_{12}+b_{32}\,\, =\,\, 0\, $,  all the other parameters being zero,
one then
 gets additional factorizations, the modified factorization scheme 
yielding :
\begin{eqnarray}
&& \rho(x) \, \,=  \, \,\, {{1} \over {1-x}}  \, , \quad \quad \quad  \quad 
{{ \beta(x) } \over {3 \, x }} \,\, = \,\, \,  {\frac {1+x^2}{1-x-{x}^{3}}}
\end{eqnarray}
and again $\lambda  \simeq 1.46557 \cdots $.
There is also {\em polynomial growth} subcases, for instance
$a_{12}=c_{22}=b_{32}=1$, $a_{11}$ and $a_{22}$ arbitrary non zero,
all the other ones being zero.
Generally speaking, having a complexity generically 
independent of $\, r \, $ parameters (here twenty one), one can only
expect {\em more} factorizations on some subvariety of the
$r$-dimensional space, and consequently a {\em smaller}
complexity value $\, \lambda$ on this very subvariety.

We now give another eleven parameter example associated
with the following linear transformation :
\begin{eqnarray}
\label{eleven22}
L : \, \quad \left [\begin {array}{ccc} 
m_{{1,1}}&m_{{1,2}}&m_{{1,3}}\\
\noalign{\medskip}m_{{2,1}}&m_{{2,2}}&m_{{2,3}}\\
\noalign{\medskip}m_{{3,1}}&m_{{3,2}}&m_{{3,3}}\end {array}\right ]
\qquad \longrightarrow \qquad 
\left [\begin {array}{ccc} m_{1,1}&m_{1,2}+b_{21}\,m_{2,1}
+ b_{22}\,m_{2,2}+b_{23}\,m_{2,3}&m_{13}\\
\noalign{\medskip}
m_{2,1}&a_{12}\,m_{1,2}+a_{21}\,m_{2,1}+a_{22}\,m_{2,2}+
a_{23}\,m_{2,3}+a_{32}\,m_{3,2}&m_{2,3}\\
\noalign{\medskip}
m_{3,1}&m_{32}+c_{21}\,m_{2,1}+c_{22}\,m_{2,2}+c_{23}\,
m_{2,3}&m_{3,3}
\end {array}\right ]
\end{eqnarray}
For $\, \,K \, = \, L \cdot I \, \,$ the corresponding generating functions
 are :
\begin{eqnarray}
&&\rho(x) \, = \, \,  \, {{1} \over {1-x}}  \, , \qquad \qquad 
{{ \beta(x) } \over {3 \, x}} \,\, = \, \, 
{\frac { 1\, -\, x^3 }{1\, - \, x\, -\, x^2 \, - \, x^3 \, + \, x^4}}
\end{eqnarray}
The numerator of $\beta(x)$ does not appear in
table Tab.~\ref{latable}: this mapping
has a {\em new value} for the complexity $\, \lambda \, \simeq \,
1.72088 \cdots $,
{\em not previously obtained} for any of the $\, 9!\, $ permutations.

Family (\ref{eleven22}), depending on eleven continuous 
parameters, also enables to 
address  the following problem :
is the complexity growth crucially dependent
on the {\em reversible character}~\cite{QuRo88}
of the transformations ?
In fact one may lose the birational character of $K$ when, 
for instance, the linear transformation $L$ 
becomes singular. This is very easy to realize for some
condition on the eleven parameters (codimension one subvariety). For instance,
taking $\,\,\, b_{22}\, =\, 2\, , \, \,  a_{22}\, =\, 87\, ,$
$ \, \, \, a_{12}\, =\, 5\, , \, \,  a_{32}\, =\, 7\, , \, \,  c_{22}\,
=\, 11\, $, all
the other parameters being zero, leads 
to a {\em non invertible} mapping 
$\, K\, =\, L \cdot I$. One easily verifies that the factorization scheme,
 the associated generating functions and thus 
the complexity $\, \lambda$, are {\em unchanged} in this case
and, more generally, on such singular subvarieties.
With this first rational, non invertible, example one sees
that the rational character of the generating functions is
not a consequence of a ``simple'' invertibility 
of the mapping (see also~\cite{BoMa95}).

\subsection{From linear transformations to homogeneous polynomial transformations}
\label{fromto}
There is nothing specific with linear transformations. For instance,
let
us consider the following quadratic transformation depending on twenty
one parameters (which is reminiscent of (\ref{linearL})) :
\begin{eqnarray}
\label{twentyone}
&&Q : \, \quad 
\left [\begin {array}{ccc} 
m_{{1,1}}&m_{{1,2}}&m_{{1,3}}\\
\noalign{\medskip}m_{{2,1}}&m_{{2,2}}&m_{{2,3}}\\
\noalign{\medskip}m_{{3,1}}&m_{{3,2}}&m_{{3,3}}
\end {array}\right ]
\qquad \longrightarrow \qquad  \\
&&\left [\begin {array}{ccc}
 m_{{1,1}}^2&a_{11}\,m_{{1,1}}^2+a_{12}\,m_{{1,2
}}^2+a_{13}\,m_{{1,3}}^2+a_{21}\,m_{{2,1}}^2+a_{22}\,m_{{2,2}}^2+a_{23}
\,m_{{2,3}}^2+a_{31}\,m_{{3,1}}^2+a_{32}\,m_{{3,2}}^2+a_{33}\,m_{{3,3}}^2&m_{{1,3}}^2\\
\noalign{\medskip}m_{{2,1}}^2&c_{21}\,m_{{2,1}}^2+c_{22}\,m_{{2,2}}^2+c_{23}\,m_{{2,3}}^2&m_{{2,3}}^2\\
\noalign{\medskip}m_{{3,1}}^2&b_{11}\,m
_{{1,1}}^2+b_{12}\,m_{{1,2}}^2+b_{13}\,m_{{1,3}}^2+b_{21}\,m_{{2,1}}^2
+ b_{22}\,m_{{2,2}}^2+b_{23}\,m_{{2,3}}^2+b_{31}\,m_{{3,1}}^2
+b_{32}\,m_{{3,2}}^2+b_{33}\,m_{{3,3}}^2&m_{{3,3}}^2
\end {array}\right ] \nonumber
\end{eqnarray}
The homogeneous transformation $K \, = \, Q \cdot I$
{\em gives again a stable factorization scheme}.
In this case, where $\, Q\, $ is no longer a linear transformation,
but a  homogeneous polynomial transformation 
of degree $r$ (here $r=2$),
the factorization scheme remains of the general
form (\ref{detgen}) and (\ref{Kgen}).
As far as generating functions are concerned 
some modifications have to be done.
Firstly the $\rho_n$ 's, and associated 
$\rho(x)$, should be replaced by the $\, \, \gamma_n$'s
defined by : 
\begin{eqnarray}
\label{defgamma}
\widehat{K}(M_n) \, \, \, = \,  \,\, \, {{ K(M_n) } \over
 {det(M_n)^r}} \,
 \, \, = \, \, \,
{{ M_{n+1} } \over {f_{n+1}^{\gamma_0} \cdot f_{n}^{\gamma_1} \cdot
 f_{n-1}^{\gamma_2}\, \cdots }}
\end{eqnarray}
and the corresponding generating function $\gamma(x)$.
The linear relations between $\eta(x)$, $\phi(x)$ and $\gamma(x)$
are slightly modified (see  (\ref{newmew0}), (\ref{gamma})
in  appendix C).
Secondly, relation (\ref{detoverdetxn}) 
is no longer valid here. A new relation has to be introduced
playing the same role.
Transformation $\, \widehat{K}\, =\, \, \, Q \cdot 
\widehat{I} \, $ is a homogeneous transformation of degree $\, -\, r$.
Instead of introducing the determinantal 
variables $\, x_n\, $ through (\ref{defxn}), let us introduce
$\tilde{x}_n$ by :
\begin{eqnarray}
\tilde{x}_n(M_0) \, \, = \, \, \, \det(\widehat{K}^{n+1}(M_0))
 \cdot \bigl(\det(\widehat{K}^{n}(M_0))\bigr)^r
\end{eqnarray}
These new determinantal variables $\tilde{x}_n$ 
 are well-suited ones since they are {\em invariant under 
a rescaling of} $M_0 \, $ : \\
$ \, \, \tilde{x}_n({\rm Cst} \cdot M_0) \, \, 
= \, \,  \tilde{x}_n(M_0)   \, $.
Relation (\ref{detoverdetxn}) becomes :
\begin{eqnarray}
\tilde{x}_n(M_0) \, \, \,  = \, \, \, \, \, \,  f_{n+1}^{W_0} 
\cdot f_n^{W_1} \cdot f_{n-1}^{W_2} 
\cdot f_{n-2}^{W_3} \,  \, \, \cdots  \, \, \,  f_0^{W_{n+1}}
\end{eqnarray}
Again one can introduce the generating function of these
exponents $W_n$ and see that relation (\ref{Wb}) 
still holds. From the stable factorization scheme of $\, K \, = \, Q \cdot I\, $
 one now gets :
\begin{eqnarray}
\label{isnot}
 {{ \beta(x)} \over {3 \, x}} \, = \, \, {{ 1}
\over { 1-\, 3 \, x \, -2 \,x^2}}\, , \, \,\quad \quad \quad
 \gamma(x) \, = \, \, \,  2 \cdot {{1-x} \over {1-2\, x}}
\end{eqnarray}
This gives a complexity value $\, \lambda\, \simeq \, 3.5615 \cdots \, $.
Let us consider the expression of $\alpha(x)$  :
\begin{eqnarray}
\label{expr}
\alpha(x) \, \, = \, \, \,
 {{ 3 \cdot (1\, +x \, -\, 2 \, x^2\, + \, 4
\, x^3 \,) }
\over { (1+\, 2 \, x) \cdot (1- \, 2 \, x) \cdot (1-\, 3 \, x \, -2
\,x^2)}}\,
\end{eqnarray}
On this expression one sees that other poles occur. 
The inverse of these additional poles, namely $\pm 2$, are actually smaller 
that the complexity value $\, 3.56155 \cdots \, $. 
The existence of ``subdominant'' poles already occurred 
with permutations of entries, or
 linear transformations (see $\, \beta(x) \, $ in~(\ref{m063})): we often had 
 $\, 1\, - \, x $, or $\, 1\, + \, x $, additional factors in the expressions
of the degree generating functions. With expression (\ref{expr})
one sees the occurrence of a $ \,\,\, 1 \, - \, 2 \cdot x\,\, $ factor
instead of $\, 1\, - \, x\, $ factors.

There is also nothing specific with quadratic transformations.
Let us  introduce the simple homogeneous polynomial
of degree $r$ :
\begin{eqnarray}
\label{Qr}
&&Q_r : \, \, \quad 
\left [\begin {array}{ccc} 
m_{{1,1}}&m_{{1,2}}&m_{{1,3}}\\
\noalign{\medskip}m_{{2,1}}&m_{{2,2}}&m_{{2,3}}\\
\noalign{\medskip}m_{{3,1}}&m_{{3,2}}&m_{{3,3}}
\end {array}\right ]
\qquad \longrightarrow \qquad 
\left [\begin {array}{ccc} 
m_{{1,1}}^r&m_{{1,2}}^r&m_{{1,3}}^r\\
\noalign{\medskip}m_{{3,2}}^r&m_{{2,2}}^r&m_{{3,1}}^r\\
\noalign{\medskip}m_{{2,3}}^r&m_{{2,1}}^r&m_{{3,3}}^r
\end {array}\right ]
\end{eqnarray}
and its associated homogeneous transformation $\, K \, = \, \, Q_r
\cdot I\, $.
Its factorization scheme is
very simple, it reads for $\, r \ge 2$ (for $\, r\, =\, 1$
transformation (\ref{Qr}), and $\, K \, = \, \, Q_r
\cdot I $, become trivial) :
\begin{eqnarray}
M_n \, = \, \, {{K(M_{n-1})} \over {f_{n-1}^r}}\, , \qquad \quad 
det(M_n) \, = \, \, \, f_{n+1} \cdot f_n^2
\end{eqnarray}
which yields the following linear relations on the $\alpha_n$'s and
$\beta_n$'s (see also appendix C) :
\begin{eqnarray}
\alpha_n \, = \, \, 2 \cdot r \cdot \alpha_{n-1} \,  \, \, - \, 3 \cdot r \cdot
\beta_{n-1}\, , \quad
\qquad \alpha_n \, = \, \,      \beta_{n+1}\,+ \, 2 \cdot \beta_{n}\,
\end{eqnarray}
It gives the following generating functions
for arbitrary $r \ge 2 $  :
\begin{eqnarray}
{{\beta(x)} \over {3 \, x}} \, 
= \, \, \,{\frac {1}{1\, + \, 2 \cdot (1-r ) \cdot x\, - \, r \cdot  x^2}}\, , \qquad \,
\, 
\eta(x) \, = \, r \, , \qquad \,
\, 
\phi(x) \, = \,1+2 \cdot x \, , \quad \quad
\gamma(x) \, = \, r \cdot (1+x)
\end{eqnarray}
For homogeneous polynomials
of degree $\, r\, $ one can show that  subdominant poles, like $\,\, 1\, - r
\cdot x $, 
may occur instead of the previous $\, 1\, -x\, $ and $ 1 -\, 2 \, x \, $ factors.

For $\, r\, = \, 2\, $, one remarks that one gets a degree generating
function :
\begin{eqnarray}
{{\beta(x)} \over {3 \, x}} \, 
= \, \, \, \,{\frac {1}{1\, -\, 2 \,x \, -\, 2 \,x^2}}
\end{eqnarray}
which is not the limit of (\ref{isnot}). 
The generic complexity corresponding to 
 (\ref{isnot}), namely $\, \lambda \,
\simeq \, 3.56155 \cdots $ is changed, for (\ref{Qr}) taken for $r\, = \,
2$, into  $\, \lambda \,
\simeq \, 2.73205 \cdots$. There actually exist many subvarieties
of the  twenty-one parameter space of transformations
(\ref{twentyone})
on which the  generic complexity 
$\, \lambda \,
\simeq \, 3.56155 \cdots $
is modified into another (smaller) 
algebraic value. One remarks that the subvarieties
of the  twenty-one parameter space of transformation
(\ref{linearL}) (for instance, $a_{12}=c_{22}=b_{32}=1$, $a_{11}$ 
and $\, a_{22}$ arbitrary non zero,
all the other ones being zero,
 previously mentioned as a polynomial growth subcase)
 also yield ``non-generic''  complexities for (\ref{twentyone}).

\section{conclusion}
In  previous papers~\cite{zeta,topo} it has been shown 
 that the topological entropy,
and the Arnold complexity, actually identify on various simple
two-dimensional birational examples, and
 that these  quantities are actually
 algebraic numbers. The generating functions corresponding to these
two complexity measures, namely the {\em dynamical zeta function}~\cite{Gu70,Ru78,Ru91}
and the various ``degree'' generating functions (like $\beta(x)$)
were shown to be simple {\em rational expressions} with integer
 coefficients~\cite{zeta}, the dominant poles in these two sets
of generating functions being the same.
When one analyzes birational transformations depending on 
more than two variables, it becomes very difficult 
to calculate even the first coefficients of the expansion 
of the dynamical zeta function. On the contrary the calculations
on the degree generating functions 
(associated to the Arnold complexity) can be quite easily
performed, even for birational transformations of many variables
(the $q^2$ entries of a matrix~\cite{BoMa95}).

Analyzing exhaustively a first finite set of  $ \, 362880 \, $ 
birational transformations (associated with 
all the permutations of $\, 3 \times 3$ matrices), 
we have obtained non-trivial, 
but still simple, ``spectrum'' of 
eighteen algebraic Arnold complexities
for the corresponding dynamical systems. In a second step
 it has been shown that 
these results can be  drastically generalized  along 
three different lines {\em preserving 
the algebraic character of the complexities}. 
Firstly, one can combine these birational transformations together,
and get extremely rich sets of  {\em algebraic} complexities. 
Secondly, one can consider (generically birational)
transformations, associated with 
linear transformations of
the entries of  $\, 3 \times 3$ matrices,
and still get  sets of algebraic complexities. 
Remarkably one has another {\em universality}
property here: these algebraic complexities do not depend
on many of the continuous parameters 
associated with the linear transformations.
Thirdly, one still gets  sets of algebraic complexities
with {\em rational} transformations 
(associated with homogeneous polynomial
 transformations on the entries) which again can depend on 
many continuous parameters.
With this last generalization we have completely
 lost any invertible character
of the transformations. On the top of that 
these $3 \times 3$ matrix calculations
 can be simply generalized to $ q \times q$
matrices  {\em for arbitrary}\footnote{The ``spectrum'' of
 values of the complexity $\, \lambda\, $  depends on $q$, 
see for instance~\cite{BoMa95}.} $q$.
Combining several of these rational transformations
depending on several continuous parameters together,
one certainly gets  again rich  sets of {\em algebraic} complexities.

We end up with an extremely large set of transformations,
so large that, clearly, it should be a powerful tool
to  study  discrete dynamical systems.

\appendix 


\section{A  ``transmutation''
 property of the matrix inversion}
\label{AppendixA1}

Let us sketch here some non trivial symmetries
between the permutations.

The transformations, considered in sections
 (\ref{factgenfunc}), (\ref{revisit}), are products 
of matrix inversion and permutations of the entries.
Any such non trivial symmetry of the birational transformations 
$\widehat{K}\, $ should correspond to a non-trivial
relation between  matrix inversion and permutations of the entries
of the matrix.
Such relations actually exist. They correspond to a ``transmutation''
property between the inversion and
 permutations $\, P$ and $\, Q$. There actually 
exist two permutations $\, P$ and $\, Q\, $ such that :
\begin{eqnarray}
\label{transmut}
   P \cdot \widehat{I} \, \,\, \, = \,\,\, \, \, \widehat{I} \cdot Q
\end{eqnarray}

Permutations, such that a ``transmutation'' relation 
(\ref{transmut}) is satisfied, do exist : one can easily build
examples by combining product of
permutations that permutes {\em only rows}
of a $q \times q\, $ matrix (that we will denote
by ``$R $''), 
permutations that permutes  {\em only columns}
of a $q \times q\, $ matrix (that we will denote
by ``$C$'') and, possibly, the matrix transposition we
denote ``$T$''. 
Examples of permutations $\, P\, $ and $\, Q$,  such that 
(\ref{transmut}) is satisfied, read :
\begin{eqnarray}
\label{decompRCT}
   P  \, \,\, \, = \,\,\, \, \, R \cdot C \cdot T^{\epsilon} \, , \,
   \qquad \hbox{where :} 
\qquad \epsilon \, = \, 0\, , \, \, \hbox{or } \, \, 1
\end{eqnarray}
and similarly for permutation $\, Q$.
A permutation $P$ having such a decomposition (\ref{decompRCT})
will be called an ``RCT'' permutation.

Let us consider two permutations $t_1$ and $t_2$,  yielding respectively
the two birational transformations
 $\,\widehat{ K}_1\, = \, t_1 \cdot \widehat{I}\, $ and
$\widehat{K}_2 \, = \, t_2 \cdot \widehat{I}\,$.
Let us introduce the following relation of equivalence between 
two permutations $t_1$ and $t_2$ :  $t_1$ and $t_2$ will be related
if they are
 such that there exists an  ``$\, R C T \, $'' permutation, 
$ b_0 $, 
such that :
\begin{eqnarray}
\label{fund}
\widehat{K}_1^n \, \,  = \, \,  \,  \,  \,   \, 
b_0 \cdot \widehat{K}_2^n \cdot b_0^{-1} \,   \, \,
\end{eqnarray}

Relation (\ref{fund}) can easily be seen to define
a relation of equivalence between $\, t_1$ and $\, t_2 $,
 we will denote  ${\cal R}^{(n)}$ :
\begin{eqnarray}
t_1 \,  \,  {\cal R}^{(n)} \, \,   t_2
\end{eqnarray}
Note  that this  $\, {\cal R}^{(n)} \, $ equivalence relation is compatible 
with the inverse in the permutation group
 $\, t \quad \longrightarrow \quad t^{-1}\, $.
Also note that  the equivalence of two permutations, 
up to simple rows and columns relabeling,
is an  $\, {\cal R}^{(1)} \, $ equivalence, however,
conversely, the  $\, {\cal R}^{(1)} \, $ equivalence
does not reduce to the simple, and quite trivial,
equivalence
of two permutations 
up to simple rows and columns 
relabeling. Obviously rows and columns relabeling 
of the matrices do not modify
their integrability properties~\cite{Zittartz}, as well as the growth of the
calculations.

It is obvious that if $\,\, \,\,  t_1 \,  {\cal R}^{(n)} \,  t_2 \,\, \,\,  $ then 
 $\,\, \,\,  t_1 \,  {\cal R}^{(n \times p)} \,  t_2 \,\,\,\,   $ 
for any natural integer $\, p\, $. This is a consequence of the fact
 that :
\begin{eqnarray}
\label{funda}
\widehat{K}_1^n \, = \, \, b_0 \cdot \widehat{K}_2^n \cdot b_0^{-1}
 \,   \, \,\quad \quad \quad
\hbox{yields :}  \quad \quad \quad \quad  \quad  \widehat{K}_1^{n\, p} \, \, =
\,  \, \, b_0 \cdot
\widehat{K}_2^{n\, p} \cdot b_0^{-1}
\end{eqnarray}
 
If two permutations, $\, t_1$ and $t_2$, are in the same 
equivalence class with respect to $\, {\cal R}^{(m)}$,
and  if $t_2$ and $\, t_3$ are in the same 
equivalence class with respect to $\, {\cal R}^{(n)}$
where $n \, \ne m\, $,  $t_1$ and $t_3$ are in the 
same equivalence class with respect to $\, {\cal R}^{(n \times m)}\, $,
or with respect to $\, {\cal R}^{(N)}\, $ for some ``large 
enough'' integer $N$.
In fact it can be shown, on 
the example of the equivalence classification
of the permutations of $\, 3 \times 3\, $ matrices, that this value of
$\, N\, $ corresponding to  the  (``asymptotic'' equivalence) relation 
is  actually equal to
 $\, N\, = \, 24\, $.

If two permutations  $\, t_1$ and $t_2$
are in the same 
equivalence class, with respect to $\, {\cal R}^{(m)}$,
the complexities (which are real positive
numbers), associated with their respective
birational transformations $\widehat{K}_1$ and $\widehat{K}_2$,
we denote $\lambda_1$ and $\lambda_2$
are, as a straight consequence of (\ref{fund}), related
by :
\begin{eqnarray}
\lambda_1^m \, = \, \, \, \lambda_2^m
\end{eqnarray}
Therefore one sees that their complexities  are equal :
 $\lambda_1 \, = \, \lambda_2$.
In particular if one considers the (largest) equivalence 
classes corresponding, for $\, 3 \times 3$ matrices, to $\, {\cal R}^{(24)}$,
all the representants in one of these $\, {\cal R}^{(24)}$
equivalence classes will have the {\em same complexity} growth $\lambda$.

\section{A ``molecular'' factorization scheme}
\label{factomol}
The factorization scheme of $ \, {\cal K} \,= \, \, t_1 \cdot I \cdot
 t_2 \cdot I$, corresponding to
 permutation $\, 146237058\, $ and  permutation $\,471562380 \, $
(see section (\ref{mol}),
is of the same type as the one described in~\cite{zeta},
namely a {\em parity-dependent} factorization scheme
(which is a straight consequence of the fact that one acts
with $K_1$, and then with $K_2$, and again ...) :
\begin{eqnarray}
\label{newnew}
&&f_1\, = \, det(M_0) \, , \quad 
M_1\,= \, K_1(M_0) \,, \quad
f_2\  = \, det(M_1) \, , \quad
M_2\,= \, \, K_2(M_1)  \, , \quad
f_3\, = \, \frac{det(M_2)}{ f_2 } \, ,\quad 
M_3\,= \, K_1(M_2)\,, \nonumber \\
&&f_4\, = \, det(M_3)\, , \qquad
M_4\,= \, K_2(M_3) \,, \qquad 
f_5\, = \, \frac{ det(M_4)}{ f_2^3
                \cdot f_4 } \, , \qquad 
M_5\,= \, \frac{K_1(M_4)}{   f_2 
    } \, , \qquad 
f_6 \, = \, \, {{det(M_5) } \over { f_2^2 \cdot f_4 }}\, ,\, \,\,
\cdots 
\end{eqnarray}
and for arbitrary $n \ge 3$ :
\begin{eqnarray}
det(M_n)\,&=& \,\,\,  f_{n+1} \cdot f_{n} \cdot f_{n-2}^3 
\cdot f_{n-6} \cdot f_{n-8} \cdot  f_{n-10}  \cdot f_{n-12}  \cdot
f_{n-14}  \,\,\, \cdots
 \nonumber \\
\label{KIValphq=3even}
K_1(M_n)\, &=&\,\,\, M_{n+1} \cdot  f_{n-2} 
\end{eqnarray}
for $n$ even and :
\begin{eqnarray}
det(M_n)\,&=&\,\,\, f_{n+1} \cdot f_{n-1}
  \cdot f_{n-3}^2 
 \cdot f_{n-5} 
\cdot f_{n-7}^2 \cdot f_{n-9}^2 \cdot
 f_{n-11}^2 \cdot f_{n-13}^2 \,\,\, \cdots \nonumber \\
\label{KIValphq=3odd}
K_2(M_n)\, &=&\,\,\, M_{n+1} \cdot  f_{n-3} \cdot f_{n-7}
\cdot f_{n-9}  \cdot f_{n-11}  \cdot f_{n-13} \,  \, \cdots 
\end{eqnarray}
for $n$ odd. This yields the following expressions for the odd and even parts of 
$\alpha(x)$ and $\beta(x)$ (``2'' for even and ``1'' for odd) :
\begin{eqnarray}
&&\beta_2(x) \, = \, \,
 {{6 \cdot x^2} \over { 1 \, -3\, x^2\,+x^4\, -\, x^6\,  -2\, x^8}}\, , \qquad
 \qquad
\beta_1(x) \, = \, \,
\,  {\frac { 3 \cdot x \cdot (1\, + \, {x}^2 )\cdot
\left (-1+x\right )^{2}\left (x+1\right )^{2}}{
1\, -3\,{x}^{2}\, +{x}^{4}\, -{x}^{6}\, -2\,{x}^{8}}}\, , \nonumber \\
&&\alpha_2(x) \, = \, \,{{  3 \cdot (1\, + 4\,{x}^{4}-4\,{x}^{6}\, +{x}^{8})}
 \over {(1-x^2) \cdot (1\, -3\,{x}^{2}\, +{x}^{4}\, -{x}^{6}\,
 -2\,{x}^{8}) }}\, , \qquad  \alpha_1(x) \,
 = \, \,{{  6\cdot x  \cdot ( 1\, +x^4\, -x^6\, +x^8)}
 \over {(1-x^2) \cdot (1\, -3\,{x}^{2}\, +{x}^{4}\, -{x}^{6}\,
 -2\,{x}^{8}) }}\,
\end{eqnarray}
These generating functions yield a ``molecular complexity'' :
$\lambda \simeq \, \,  2.8581 \cdots $.
These generating functions verify a parity dependent
 system of functional relations
 which generalizes the ones described in~\cite{BoMa95} :
\begin{eqnarray}
&&x \cdot \alpha_1(x) \, - \, \beta_2(x) \, = \, \, F_{2p}(x) \cdot
\beta_2(x)\, , \qquad  \qquad  \qquad 
 x \cdot \alpha_2(x) \, - \, \beta_1(x) \, = \, \, F_{1m}(x) \cdot \beta_2(x)\,
\, ,  \\
&&\alpha_2(x)\, - \, 3 \, -\, 2 \cdot x \cdot \alpha_1(x)\,
+ \, 3 \cdot G_{2p} \cdot \beta_2(x)\,\,= \, \, 0
 , \qquad \alpha_1(x)\, \, -\, 2 \cdot x \cdot \alpha_2(x)\,
+ \, 3 \cdot G_{1m} \cdot \beta_2(x)\,= \, \, 0 \nonumber
\end{eqnarray}
where :
\begin{eqnarray}
&&F_{2p}(x) \, = \, \, {x}^{2}+2\,{x}^{4}+{x}^{6}\, +\, \,{\frac
{2 \cdot {x}^{8}}{1-{x}^{2}}}\, , \, \quad 
F_{1m} \, = \, \, 2\,{x}^{3}-{x}^{5}+{\frac {x}{1-{x}^{2}}}\, , \,
\quad \,   G_{1m}(x)\, = \, \, x^3\,,   \, \, \quad  G_{2p} \, = \, \,
{x}^{4}+{\frac {{x}^{8}}{1-{x}^{2}}}
\nonumber
\end{eqnarray}

\section{Exponent generating functions for homogeneous
polynomial transformations of degree $r$}
Let us consider a homogeneous transformation $\, Q_r\, $ of degree
$r$ (like (\ref{Qr}), or like (\ref{twentyone}) for $\, r=2$) 
and its associated  homogeneous transformation 
$\, K \, = \, Q_r \cdot I\, $.
Relations (\ref{detgen}), (\ref{Kgen}) are still valid 
but yield a slight modification of the linear functional 
relations  (\ref{etax}) and (\ref{defzet}) , namely :
\begin{eqnarray}
\label{newmew0}
&&((q-1 ) \cdot r \cdot x\, -\, 1 ) \cdot \alpha(x) \, \, + \, q \,
  -\,  q \cdot x \cdot \eta(x) \cdot  \beta(x) 
  \,\, \, = \, \,\, \, 0 \,, \\
\label{newmew01}
&&x \cdot \alpha(x) \, \,  \,  = \, \, \,   \, \phi(x) \cdot \beta(x)
\end{eqnarray}

Let us recall that, for homogeneous transformations of degree
$r$, one must introduce, 
instead of $\, \rho(x)$, the generating
function $\, \gamma(x)\, $  (see section (\ref{fromto})) 
defined by :
\begin{eqnarray}
\widehat{K}(M_n) \, = \, \, {{ K(M_n) } \over {det(M_n)^r}} \, = \, \,
{{ M_{n+1} } \over {f_{n+1}^{\gamma_0} \cdot f_{n}^{\gamma_1} 
\cdot f_{n-1}^{\gamma_2}\cdots }}
\end{eqnarray}
This last relation yields a new relation :
\begin{eqnarray}
\label{gamma}
 q \, + \, q\cdot  \gamma(x) \cdot  \beta(x) \, \, = \, \, \,  (1 \, + \, r
  \cdot x) \cdot  \alpha(x)
\end{eqnarray}
which has to be compatible with the previous 
two (\ref{newmew0}), (\ref{newmew01}) :
\begin{eqnarray}
r \cdot \phi(x) \, \, = \, \, \, \,
 \gamma(x) \,  \, + \, x \cdot \eta(x) 
\end{eqnarray}



\begin{references}
\bibitem{C}
	The multi-precision library gmp is part of the GNU project.
\end{references}
\end{document}